\documentclass[preprint2]{aastex}

\usepackage{color}

\shorttitle{LATITUDE DEPENDENCE OF DIFFUSE GALACTIC LIGHT}

\shortauthors{Sano et al.}

\begin{document}


%\title{NEAR-INFRARED SCATTERING PROPERTIES OF INTERSTELLAR DUST GRAINS REVEALED BY DIRBE REANALYSIS}

%\title{NEAR-INFRARED STRONG FORWARD SCATTERING CHARACTERISTIC OF INTERSTELLAR DUST GRAINS REVEALED BY DIRBE REANALYSIS}
\title{FIRST DETECTION OF GALACTIC LATITUDE DEPENDENCE OF NEAR-INFRARED DIFFUSE GALACTIC LIGHT FROM DIRBE REANALYSIS}


\author{K. SANO\altaffilmark{1,2},
 %K. KAWARA\altaffilmark{3}, 
 S. MATSUURA\altaffilmark{3}, 
 K. TSUMURA\altaffilmark{4},\\
 %H. KATAZA\altaffilmark{1,2}, 
T. ARAI\altaffilmark{4}, 
M. SHIRAHATA\altaffilmark{4}, 
Y. ONISHI\altaffilmark{2,5}}
%Y. MATSUOKA\altaffilmark{6}}

\affil{
\altaffilmark{1}Department of Astronomy, Graduate School of Science, The University of Tokyo, \\
Hongo 7-3-1, Bunkyo-ku, Tokyo 113-0033, Japan\\
\altaffilmark{2}Institute of Space and Astronautical Science, Japan Aerospace Exploration Agency,\\
3-1-1 Yoshinodai, Chuo-ku, Sagamihara, Kanagawa 252-5210, Japan\\
\altaffilmark{3}Department of Physics, School of Science and Engineering, Kwansei Gakuin University, 2-1 Gakuen, Sanda, Hyogo 669-1337, Japan \\
\altaffilmark{4}Frontier Research Institute for Interdisciplinary Science, Tohoku University, Sendai 980-8578, Japan \\
\altaffilmark{5}Department of Physics, Tokyo Institute of Technology 2-12-1 Ookayama, Meguro-ku, Tokyo, 152-8550, Japan \\
}

\email{sano@ir.isas.jaxa.jp}

%\and

%\author{R. J. Hanisch\altaffilmark{5}}
%\affil{Space Telescope Science Institute, Baltimore, MD 21218}


%\altaffiltext{1}{Visiting Astronomer, Cerro Tololo Inter-American Observatory.
%CTIO is operated by AURA, Inc.\ under contract to the National Science
%\altaffiltext{2}{Society of Fellows, Harvard University.}
%\altaffiltext{3}{present address: Center for Astrophysics,
 %   60 Garden Street, Cambridge, MA 02138}
%\altaffiltext{4}{Visiting Programmer, Space Telescope Science Institute}
%\altaffiltext{5}{Patron, Alonso's Bar and Grill}


\begin{abstract}
Observational study on near-infrared (IR) scattering properties of interstellar dust grains has been limited due to its faintness.
Using all-sky maps obtained from Diffuse Infrared Background Experiment (DIRBE), we investigate the scattering property from diffuse Galactic light (DGL) measurements at $1.25$, $2.2$, and $3.5\,\rm{\mu m}$ in addition to our recent analyses of diffuse near-IR emission (Sano et al. 2015; Sano et al. 2016).
As a result, we first find that the intensity ratios of near-IR DGL to $100\,\rm{\mu m}$ emission increase toward low Galactic latitudes at $1.25$ and $2.2\,\rm{\mu m}$.
The derived latitude dependence can be reproduced by a scattered light model of interstellar dust with a large scattering asymmetry factor $g\equiv\langle\cos\theta\rangle$ of $0.8^{+0.2}_{-0.3}$ at $1.25$ and $2.2\,\rm{\mu m}$, assuming an infinite Galaxy disk as an illuminating source.
The derived asymmetry factor is comparable to the values obtained in the optical, but several times larger than that expected from a recent dust model.
Since possible latitude dependence of ultraviolet-excited dust emission at $1.25$ and $2.2\,\rm{\mu m}$ would reduce the large asymmetry factor to the reasonable value, our result may indicate the first detection of such an additional emission component in the diffuse interstellar medium.

\end{abstract}





\section{INTRODUCTION}

Various astrophysical phenomena in the universe are greatly affected by properties of interstellar dust grains, such as the albedo, scattering asymmetry, composition, and size distribution. 
For example,  these properties determine the interstellar extinction curve.
At ultraviolet (UV), optical, and near-infrared (IR) wavelengths, the scattering properties of the dust grains can be investigated by scattered light measurements.
In fact, such studies have been conducted at UV and optical wavelengths (e.g., Lillie \& Witt 1976; Schiminovich et al. 2001; Witt et al. 1997).
In the near-IR, however, scattered light measurements have been limited due to the low optical depth.
By surface brightness observations of a globule, Lehtinen \& Mattila (1996) derived the near-IR grain albedo, but no study has determined the near-IR scattering asymmetry factor which represents the degree of forward scattering.
Particularly in the near-IR, isotropic scattering due to the Rayleigh scattering is expected to dominate because typical grain size is thought to be much smaller than the wavelength in recent dust models (e.g., Weingartner \& Draine 2001; Zubko et al. 2004; Compi\`egne et al. 2011).
Therefore, determination of the near-IR scattering asymmetry factor is crucial in constraining the size distribution of the interstellar dust grains.

Diffuse Galactic light (DGL) consists of the scattered light and thermal emission from interstellar dust grains which are illuminated by interstellar radiation field (ISRF).
Therefore, the DGL measurement is useful to investigate the properties of the interstellar dust.
Additionally, in measurements of optical to near-IR extragalactic background light (EBL), the DGL must be subtracted correctly along with the other foreground emissions: zodiacal light (ZL) and integrated starlight (ISL).
In the DGL measurement, interstellar far-IR $100\,\rm{\mu m}$ emission is suggested as an appropriate tracer of the DGL (e.g., Brandt \& Draine 2012).
At optical wavelengths, Witt et al. (2008), Matsuoka et al. (2011), and Brandt \& Draine (2012) have reported a linear correlation between the DGL and $100\,\rm{\mu m}$ emission at high Galactic latitude regions.
In contrast, the near-IR DGL measurement has been difficult because of its faintness along with the ZL and ISL contamination (Arendt et al. 1998; Matsumoto et al. 2015).
Recently, Tsumura et al. (2013) and Arai et al. (2015) found the near-IR DGL component in the analysis of the data obtained from {\it AKARI} and Cosmic Infrared Background Experiment (CIBER), respectively, though the analyzed regions are limited in the sky.
Sano et al. (2015, hereafter Paper I) succeeded in detecting the DGL component at $1.25$ and $2.2\,\rm{\mu m}$ at general high Galactic latitudes, reanalyzing all-sky maps obtained from Diffuse Infrared Background Experiment (DIRBE) with more precise ISL evaluation by Two Micron All-Sky Survey (2MASS) point source catalog (Cutri et al. 2003; Skrutskie et al. 2006). 
In addition, Sano et al. (2016, hereafter Paper II) reanalyzed the DIRBE data at $3.5$ and $4.9\,\rm{\mu m}$ with the Wide-field Infrared Survey Explorer (WISE) catalog (Wright et al. 2010) for the ISL evaluation and found the DGL component at high Galactic latitudes.
Compared with a model of scattered light (Brandt \& Draine 2012) and dust emission (Draine \& Li 2007), Paper II indicates that the scattered light dominates the DGL at shorter near-IR wavelengths ($\lambda \lesssim2\,\rm{\mu m}$).

In Paper I and II, we derived the intensity ratios of near-IR DGL to $100\,\rm{\mu m}$ emission from the entire data at high Galactic latitudes ($|b|>35^{\circ}$), assuming the ratios are invariant throughout the sky.
However, as described in subsection 3.1 of this paper, the ratios are expected to increase toward low latitudes if the interstellar dust grains have the forward scattering characteristic (Jura 1979).
If such Galactic latitude dependence is found observationally, the scattering asymmetry factor can be derived in comparison with a scattered light model.
Thanks to its all-sky coverage with high sensitivity for the diffuse radiation, we expect that only the DIRBE data enable us to find the latitude dependence determined by the asymmetry factor of the grains.
Using the DIRBE data, we find that the ratios increase toward low latitudes at $1.25$ and $2.2\,\rm{\mu m}$ and that the latitude dependence is reproduced by a scattered light model with a large asymmetry factor of $0.8^{+0.2}_{-0.3}$ in these bands.
This value is several times larger than that expected from a recent dust model, indicating the strong forward scattering.
However, the large asymmetry factor is in conflict with various observations of interstellar dust.
If additional near-IR emission from UV-excited dust exists in the diffuse interstellar medium, the large asymmetry factor may become smaller since such an emission component is also expected to contribute to the derived latitude dependence. 
The near-IR DGL results obtained in Paper I and II are supplemented by the present study.

The remainder of this paper is as follows.
In Section 2, we briefly describe the analysis conducted in this study and present the obtained results.
In Section 3, we derive the scattering asymmetry factor assuming a simple scattered light model.
We also compare the derived asymmetry factor with previous studies in the optical and a prediction from a recent dust model.
Summary appears in Section 4.

%% In a manner similar to \objectname authors can provide links to dataset
%% hosted at participating data centers via the \dataset{} command.  The
%% second curly bracket argument is printed in the text while the first
%% parentheses argument serves as the valid data set identifier.  Large
%% lists of data set are best provided in a table (see Table 3 for an example).
%% Valid data set identifiers should be obtained from the data center that
%% is currently hosting the data.
%%
%% Note that AASTeX interprets everything between the curly braces in the 
%% macro as regular text, so any special characters, e.g. "#" or "_," must be 
%% preceded by a backslash. Otherwise, you will get a LaTeX error when you 
%% compile your manuscript.  Special characters do not 
%% need to be escaped in the optional, square-bracket argument.

%% In this section, we use  the \subsection command to set off
%% a subsection.  \footnote is used to insert a footnote to the text.


%% This section also includes several of the displayed math environments
%% mentioned in the Author Guide.

\section{ANALYSIS AND RESULT}

\begin{table*}
\begin{center}
 \renewcommand{\arraystretch}{0.9}
 \caption{Fitting results at different Galactic latitude regions in each band}
  \label{symbols}
  \scalebox{0.9}{
  \begin{tabular}{lccccc}
  \hline
  \hline
 
   Band ($\rm{\mu m}$) & Region (deg) & Number of pixels & $\nu_ib_i\,(\rm{nWm^{-2}sr^{-1}/MJy\,sr^{-1}})$ & $c_i\,(\rm{Dimensionless})$ & $\nu_id_i\,(\rm{nWm^{-2}sr^{-1}})$\\
   \hline
   \hline
   $1.25$ & $|b|>35$ & $116578$ & $4.72\pm0.02$ & $1.0236\pm0.0003$ & $60.66\pm0.08$\\
   \hline
   $1.25$ & $25<|b|<30$ & $15683$ & $6.64\pm0.03$ & $1.0582\pm0.0008$ & $59.89\pm0.26$\\
   $1.25$ & $30<|b|<35$ & $17108$ & $6.08\pm0.03$ & $1.0356\pm0.0008$ & $63.77\pm0.22$\\
   $1.25$ & $35<|b|<40$ & $17215$ & $4.33\pm0.03$ & $1.0192\pm0.0008$ & $67.06\pm0.20$\\
   $1.25$ & $40<|b|<45$ & $17136$ & $3.98\pm0.04$ & $1.0086\pm0.0009$ & $66.60\pm0.18$\\
   $1.25$ & $45<|b|<90$ & $82227$ & $3.26\pm0.03$ & $1.0063\pm0.0004$ & $62.57\pm0.06$\\
   \hline
   \hline
   $2.2$ & $|b|>35$ & $119394$ & $1.46\pm0.01$ & $1.0330\pm0.0004$ & $27.69\pm0.04$\\
   \hline
   $2.2$ & $25<|b|<30$ & $15772$ & $1.80\pm0.02$ & $1.0725\pm0.0010$ & $29.88\pm0.12$\\
   $2.2$ & $30<|b|<35$ & $17278$ & $1.62\pm0.02$ & $1.0444\pm0.0010$ & $30.53\pm0.10$\\
   $2.2$ & $35<|b|<40$ & $17167$ & $1.14\pm0.02$ & $1.0315\pm0.0010$ & $30.44\pm0.10$\\
   $2.2$ & $40<|b|<45$ & $17303$ & $1.01\pm0.02$ & $1.0204\pm0.0011$ & $30.17\pm0.09$\\
   $2.2$ & $45<|b|<90$ & $84924$ & $1.06\pm0.02$ & $1.0190\pm0.0005$ & $28.11\pm0.03$\\
   \hline
   \hline
   $3.5$ & $|b|>35$ & $103709$ & $1.20\pm0.01$ & $0.8992\pm0.0010$ & $8.92\pm0.04$\\
   \hline
   $3.5$ & $25<|b|<30$ & $12541$ & $1.22\pm0.02$ & $0.9406\pm0.0023$ & $9.56\pm0.09$\\
   $3.5$ & $30<|b|<35$ & $14279$ & $1.15\pm0.02$ & $0.9116\pm0.0023$ & $10.28\pm0.08$\\
   $3.5$ & $35<|b|<40$ & $13965$ & $0.94\pm0.02$ & $0.8751\pm0.0026$ & $11.04\pm0.08$\\
   $3.5$ & $40<|b|<45$ & $13093$ & $0.97\pm0.02$ & $0.8690\pm0.0029$ & $10.58\pm0.07$\\
   $3.5$ & $45<|b|<90$ & $75366$ & $1.06\pm0.01$ & $0.8672\pm0.0013$ & $9.33\pm0.03$\\
    \hline
    \end{tabular}
    }
    \end{center}
    \medskip

    Note. --- The parameters $b_i$, $c_i$, and $d_i$ are defined in Equation (2).
    These results are derived with the ZL coefficient $a_i$ fixed to the value determined in the $|b|>35^{\circ}$ region, i.e., $1.0079\pm0.0001$, $1.0447\pm0.0002$, and $1.1531\pm0.0003$ at $1.25$, $2.2$, and $3.5\,\rm{\mu m}$, respectively.
    Error associated with each parameter represents the statistical uncertainty determined by the fitting.

 \end{table*}
 
To derive the intensity ratios of near-IR DGL to $100\,\rm{\mu m}$ emission, we conduct the analysis that decompose the DIRBE sky brightness into the ZL, DGL, ISL, and residual emission in the same way as in Paper I and II.
We briefly describe an outline of the analysis.

For the DIRBE data, we use the solar elongation $(\epsilon) = 90^\circ$ maps in which the value at each pixel is interpolated to represent the sky brightness when $\epsilon$ is close to $90^\circ$ (COBE/DIRBE explanatory supplement 1998).
The DIRBE products are available at the website,\\ ``lambda.gsfc.nasa.gov/product/cobe/''.
Owing to the difficulty in evaluating the ZL component at $4.9\,\rm{\mu m}$ (Paper II), we analyze the other three near-IR DIRBE bands, i.e., $1.25$, $2.2$, and $3.5\,\rm{\mu m}$.

If we define $I_i({\rm Model})$ as a brightness model for the DIRBE intensity $I_i({\rm Obs})$ in the $i$ band ($i = 1.25$, $2.2$, or $3.5\,\rm{\mu m}$), we describe $I_i({\rm Model})$ as a linear combination of the intensity of the ZL, DGL, ISL, and residual emission, denoted as $I_i({\rm ZL})$, $I_i({\rm DGL})$, $I_i({\rm ISL})$, and $I_i({\rm Resid})$ in units of ${\rm MJy\,sr^{-1}}$, respectively:
 \begin{equation}
{\it I_i({\rm Model})=I_i({\rm ZL})+I_i({\rm DGL})+I_i({\rm ISL})+I_i({\rm Resid})}
\end{equation}
\begin{equation}
= {\it a_iI_i({\rm Kel})+b_i(I_{{\rm SFD}}-{\rm 0.78\,MJy\,sr^{-1}})+c_iI_i({\rm DISL})+d_i},
\end{equation}
where $I_i({\rm Kel})$, $I_{{\rm SFD}}$, and $I_i({\rm DISL})$ represent the intensity of the DIRBE ZL model (Kelsall et al. 1998), diffuse $100\,\rm{\mu m}$ emission (Schlegel et al. 1998), and the 2MASS or WISE-derived ISL map, respectively.
\footnote{The ISL maps created in Paper I and II are available in the electronic edition of the {\it Astrophysical Journal}.}
The value ${\rm 0.78\,MJy\,sr^{-1}}$ corresponds to the EBL intensity at $100\,\rm{\mu m}$ (Lagache et al. 2000).
After excluding regions around bright sources, we determine the parameters $a_i$, $b_i$, $c_i$, and $d_i$ by a $\chi^2$ minimum analysis between $I_i({\rm Obs})$ and $I_i({\rm Model})$.
\footnote{The term $I_i({\rm DGL})$ is defined in different way between Paper I and II.
Because the present analysis adopts the definition in Paper II, the fitting results at  $1.25$ and $2.2\,\rm{\mu m}$ slightly differ from those presented in Paper I.}

In the present analysis, the parameter $a_i$ is fixed to the all-sky value ($|{\it b}| > 35^\circ$) in each band to reduce the deviation caused by the brightest ZL component.
We then conduct the $\chi^2$ minimum analysis in five different Galactic latitude regions, i.e., $25^{\circ}<|b|<30^{\circ}$, $30^{\circ}<|b|<35^{\circ}$, $35^{\circ}<|b|<40^{\circ}$, $40^{\circ}<|b|<45^{\circ}$, and $45^{\circ}<|b|<90^{\circ}$.
Due to the faintness of the DGL component in the high Galactic latitudes, we analyze the wider field ($45^{\circ}<|b|<90^{\circ}$) there.
The parameters $b_i$, $c_i$, and $d_i$ determined in each region are listed in Table 1.
At $1.25$ and $2.2\,\rm{\mu m}$, we find that the intensity ratios of near-IR DGL to $100\,\rm{\mu m}$ emission, $\nu_i b_i = [3000/\lambda_i (\rm{\mu m})] {\it b_i} \,(\rm{nWm^{-2}sr^{-1}/MJy\,sr^{-1}})$ increase toward low Galactic latitudes.
In the optical wavelength, such trend may be seen in compilation of several results obtained in different fields of the sky (see Figure 10 of Ienaka et al. 2013).
Similar to the parameter $\nu_i b_i$, the ISL coefficient $c_i$ increases toward the low Galactic latitudes.
As discussed in Paper I and II, this phenomenon may be caused if faint stars not on the catalog increase toward low Galactic latitudes due to the mask by bright sources and/or the  spatial distribution of the intensity of bright and faint stars changes in different regions.

In each band, the parameters $\nu_ib_i$ and $c_i$ at $|{\it b}| > 35^\circ$ are not the intermediate value of those at $35^{\circ}<|b|<40^{\circ}$, $40^{\circ}<|b|<45^{\circ}$, and $45^{\circ}<|b|<90^{\circ}$, but larger than those results (Table 1).
Because the DGL and ISL intensity is naturally expected to change as a function of Galactic latitude, the parameters $\nu_ib_i$ and $c_i$ might be biased to larger values when analyzing a field of the wide range in Galactic latitude, e.g., $|{\it b}| > 35^\circ$.
This effect may cause the residual emission $\nu_id_i$ at $|{\it b}| > 35^\circ$ smaller than those obtained in the divided regions.
However, these differences do not change the conclusion in Paper I and II because the parameter variation among each divided region is within the total uncertainty of the results in the two papers.   
The discussion on the derived isotropic residual emission $\nu_id_i$ including the near-IR EBL is described in Paper I and II.

%% The equation environment wil produce a numbered display equation.

%% The \notetoeditor{TEXT} command allows the author to communicate
%% information to the copy editor.  This information will appear as a
%% footnote on the printed copy for the manuscript style file.  Nothing will
%% appear on the printed copy if the preprint or
%% preprint2 style files are used.

%% The eqnarray environment produces multi-line display math. The end of
%% each line is marked with a \\. Lines will be numbered unless the \\
%% is preceded by a \nonumber command.
%% Alignment points are marked by ampersands (&). There should be two
%% ampersands (&) per line.

%% Putting eqnarrays or equations inside the mathletters environment groups
%% the enclosed equations by letter. For instance, the eqnarray below, instead
%% of being numbered, say, (4) and (5), would be numbered (4a) and (4b).
%% LaTeX the paper and look at the output to see the results.

%% This section contains more display math examples, including unnumbered
%% equations (displaymath environment). The last paragraph includes some
%% examples of in-line math featuring a couple of the AASTeX symbol macros.

\section{DISCUSSION}

We discuss the derived Galactic latitude dependence of the intensity ratios of DGL to $100\,\rm{\mu m}$ emission in comparison with a scattered light model in the Milky Way.
In Figure 1, we plot the derived parameters $\nu_i b_i$ in each band as a function of Galactic latitude.
Here, vertical error bar associated with each value denotes the standard deviation of the parameters $\nu_i b_i$ determined at six longitude-divided regions in each latitude, i.e., $0^{\circ}<l<60^{\circ}$, $60^{\circ}<l<120^{\circ}$, $120^{\circ}<l<180^{\circ}$, $180^{\circ}<l<240^{\circ}$, $240^{\circ}<l<300^{\circ}$, and $300^{\circ}<l<360^{\circ}$. 
This uncertainty estimation taking into account the regional variation of the parameter is identical to that adopted in the analysis of all-sky $|{\it b}| > 35^\circ$ region (Paper I and II).

\subsection{Derivation of the Scattering Asymmetry Factor}

\begin{figure*}
\begin{center}
 \includegraphics[scale=0.8]{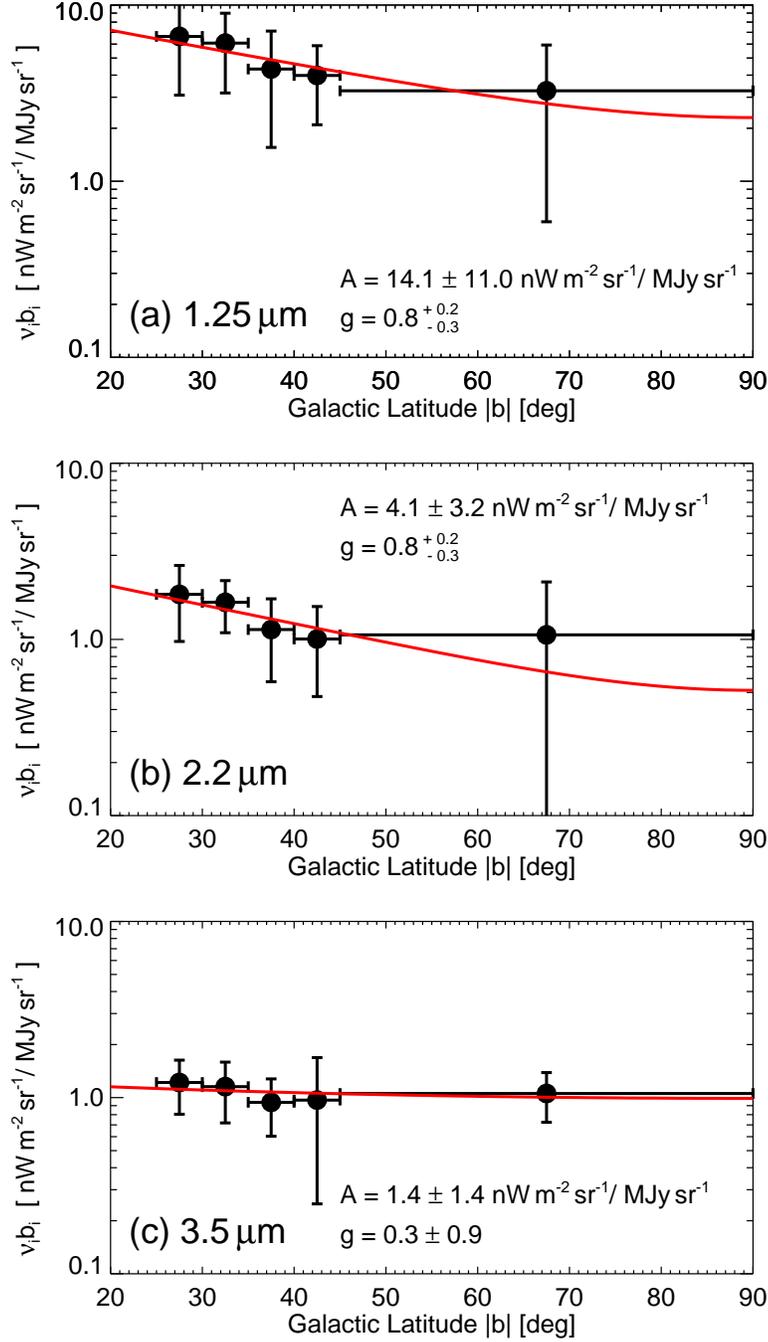} 
 \caption
 {Intensity ratios of the DGL to $100\,\rm{\mu m}$ emission $\nu_i b_i = [3000/\lambda_i (\rm{\mu m})] {\it b_i} \,(\rm{nWm^{-2}sr^{-1}/MJy\,sr^{-1}})$ as a function of Galactic latitude $|b|$.
Panels (a), (b), and (c) represent the results at $1.25$, $2.2$, and $3.5\,\rm{\mu m}$, respectively.
 Filled circle indicates the value $\nu_i b_i$ determined at each field with horizontal error bar denoting the Galactic latitude range of the analyzed field and vertical error bar the standard deviation of the parameter $\nu_i b_i$ obtained in six longitude-divided regions in each latitude.
In each panel, red curve represents the modeled intensity ratio of the scattered light (Jura 1979) to far-IR emission (Ienaka et al. 2013) in a form of $A \Bigl(1-1.1g\sqrt{\sin|b|}\Bigr)$, fitted to the derived values $\nu_i b_i$.
 The determined values $A$ and $g$ are also described in each panel.}
\end{center}
\end{figure*}

Adopting the Henyey \& Greenstein (1941) phase function for interstellar scattering, Jura (1979) numerically calculated the scattered light intensity $I_{\rm sca}$ toward the region of optical depth $\tau$ and Galactic latitude $|b|$, illuminated by an infinite homogeneous disk.
Bernstein et al. (2002) practically rewrote the intensity $I_{\rm sca}$ as
\begin{equation}
I_{\rm sca} = I_{\rm ISRF} \omega \tau \Bigr(1-1.1g\sqrt{\sin|b|}\Bigr),
\end{equation}
where $I_{\rm ISRF}$ and $\omega$ denote the ISRF intensity in the solar neighborhood (e.g., Mathis et al. 1983) and grain albedo, respectively.
The scattering asymmetry factor $g$ is defined as $g\equiv\langle\cos\theta\rangle$ where $\theta$ denotes the scattering angle from the forward direction, indicating $g=0$ for the fairly isotropic scattering and $g=1$ for the completely forward scattering.
In Equation (3), the parameters $\omega$ and $g$ are assumed as constant throughout the sky.

As a solution of simple radiative transfer of starlight and scattered light through a dusty slab (see Appendix of Ienaka et al. 2013), the intensity of far-IR emission $I_{\rm FIR}$ is expressed as
\begin{equation}
I_{\rm FIR} \propto I_{\rm ISRF} \bigl[1-\exp \{-(1-\omega)\tau\}\bigr].
\end{equation}
As shown in Figure 11 of Brandt \& Draine (2012), the optical depth in the V band is less than $\sim 0.2$ in most of the high Galactic latitudes $|b|\gtrsim 20^{\circ}$, assuming $R_V = 3.1$ Milky Way dust.
In the near-IR high latitude region of our interest, $I_{\rm FIR}$ can be approximated as
\begin{equation}
I_{\rm FIR} \propto I_{\rm ISRF} (1-\omega)\tau.
\end{equation}
From Equation (3) and (5), the intensity ratio of the scattered light to far-IR emission is modeled as
\begin{equation}
I_{\rm sca}/I_{\rm FIR} \propto \frac{\omega}{1-\omega} \Bigl(1-1.1g\sqrt{\sin|b|}\Bigr).
\end{equation}
According to this formula, in case of fairly isotropic scattering ($g=0$) the intensity ratio $I_{\rm sca}/I_{\rm FIR}$ is insensitive to Galactic latitude.
Conversely, the value $I_{\rm sca}/I_{\rm FIR}$ increases toward low latitudes in case of forward scattering ($0<g\leq1$).

To determine the scattering asymmetry factor $g$ in each band, the values $\nu_i b_i$ obtained at the five different Galactic latitude regions (Figure 1) are fitted to the following function:
\begin{equation}
\nu_i b_i = A \Bigl(1-1.1g\sqrt{\sin|b|}\Bigr),
\end{equation}
where $A$ and $g$ are the free parameters.
The result is indicated by red curve in each panel of Figure 1.
The derived value of $g = 0.8^{+0.2}_{-0.3}$ at $1.25$ and $2.2\,\rm{\mu m}$ indicates the strong forward scattering characteristic of the dust grains.
In contrast, the $g$-value is not significantly determined at $3.5\,\rm{\mu m}$ with the large uncertainty.
This can be reasonably explained by the situation that the thermal emission component in the DGL is comparable to or larger than the scattered light at $3.5\,\rm{\mu m}$.
In fact, this situation is indicated in Figure 5 of Paper II, in which the DGL result at $3.5\,\rm{\mu m}$ is compared with the model of the scattered light (Brandt \& Draine 2012) and thermal emission (Draine \& Li 2007).

\subsection{Interpretation of the Derived Galactic Latitude Dependence}

\begin{figure*}
\begin{center}
 \includegraphics[scale=0.65]{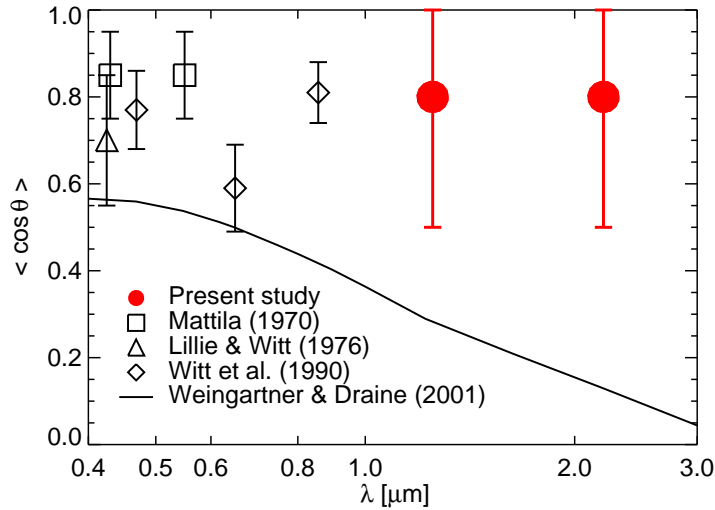} 
 \caption
 {Optical to near-IR scattering asymmetry factor $g\equiv\langle\cos\theta\rangle$.
 The present results at $1.25$ and $2.2\,\rm{\mu m}$ are indicated by filled red circles.
 The triangle, diamonds, and squares denote the results derived from Lillie \& Witt (1976), Witt et al. (1990), and Mattila et al. (1970), respectively.
 Solid curve represents the value expected from the WD01 model with $R_V=3.1$ Milky Way dust (Draine 2003).}
\end{center}
\end{figure*}

In Figure 2, we compare the derived asymmetry factor $g$ with the optical results toward some clouds (Mattila 1970; Witt et al. 1990) or a general interstellar field (Lillie \& Witt 1976).
Black solid line indicates the prediction from a recent dust model composed of carbonaceous, silicate, and polycyclic aromatic hydrocarbon (PAH) grains (Weingartner \& Draine 2001, hereafter WD01) with $R_V=3.1$ Milky Way dust (Draine 2003).
The WD01 model reportedly reproduces the observed interstellar extinction curve from UV to near-IR (Fitzpatrick 1999; Draine 2001).
The optical results prefer the strong forward scattering with $g\gtrsim 0.6$, comparable to the present near-IR values.
In the near-IR, the present results can be several times larger than the WD01 prediction.
Simply, the discrepancy can be explained if the larger dust grains are more dominant compared to the WD01 model.
The WD01 value decreases toward the longer wavelengths because the isotropic scattering by the Rayleigh scattering becomes more dominant.
In the WD01 model, typical grain size is $\sim 0.12\,\rm{\mu m}$ (Draine 2001), which is much smaller than the near-IR wavelengths.
To explain the derived large $g$-value, a lot of large grains may be required to cancel out the isotropic scattering by the smaller grains.
However, the existence of a lot of large grains is expected to conflict with various observed properties of interstellar dust, such as interstellar depletion and wavelength dependence of interstellar extinction and polarization.
This may indicate that the actual near-IR $g$-value is smaller than that obtained in the present study.

When interstellar dust grains are exposed to UV photons, continuous near-IR emission is created as a result of the stochastic heating of the small grains and/or fluorescence by large molecules such as PAH.
In fact, the UV-excited dust emission component has been observed in reflection nebulae at $1.25$--$2.2\,\rm{\mu m}$ wavelengths (Sellgren et al. 1992; Sellgren et al. 1996).
If such an emission component also exists in the general diffuse interstellar medium, the UV-excited dust emission is expected to be higher toward the low Galactic latitudes because from the {\it Galaxy Evolution Explorer} ({\it GALEX}) data, near-UV to far-UV ratio in the low Galactic latitudes is reported to be smaller than that in the high latitudes by more than a factor of two (Figure 10 of Murthy 2014).
The possible Galactic latitude dependence of the UV-excited dust emission would also contribute to the derived latitude dependence and may reduce the very large asymmetry factor $g$ to the reasonable value.
In this sense, the derived latitude dependence in the present study may indicate the first detection of UV-excited dust emission in the diffuse interstellar medium at $1.25$ and $2.2\,\rm{\mu m}$.
Similarly, to explain very high albedo derived in the galaxy NGC 4826 in the $K'$-band, Witt et al. (1994) suggested the contribution from the near-IR dust emission in addition to the scattered light.
To determine the relative contribution of the scattered light and UV-excited dust emission from the derived latitude dependence, detailed radiative transfer analysis will be needed.

The dust temperature could influence the ratio $I_{\rm sca}/I_{\rm FIR}$.
In some dark clouds, emissivity of the far-IR emission has been reported to decrease as the temperature increases (e.g., Lehtinen et al. 2007).
However, the temperature is nearly invariant ($\sim18\,{\rm K}$) throughout the general high Galactic latitude region of our interest (Schlegel et al. 1998).
This indicates that the dust temperature does not contribute to the obtained Galactic latitude dependence of $I_{\rm sca}/I_{\rm FIR}$.

According to Formula (6), if the grain albedo increases toward low Galactic latitudes, smaller $g$-value may be enough to explain the derived latitude dependence.
Notably, the modeled intensity ratio $I_{\rm sca}/I_{\rm FIR}$ is relatively sensitive to the albedo.
Lehtinen \& Mattila (1996) derived the near-IR grain albedo of $0.57<\omega<0.80$ and $0.46<\omega<0.76$ at $J$ and $K$ band, respectively.
For example, if the albedo is changed from $0.5$ to $0.7$ in Formula (6), the intensity ratio is approximately doubled.
However, such variation in the albedo is probably unrealistic because the analyzed region is limited to the general high Galactic latitudes ($|b|\gtrsim 20^{\circ}$) in the present study.
In addition, even if the weak dependence on the Galactic latitude at $3.5\,\rm{\mu m}$ (panel (c) of Figure 1) is entirely caused by the latitude dependence of the albedo, such small variation does not create the stronger latitude dependence at $1.25$ and $2.2\,\rm{\mu m}$ (panels (a) and (b) of Figure 1).

In the scattered light model (Equation (3)), Jura (1979) adopted the simplest Milky Way model in which the dust grains are illuminated by the infinite homogeneous disk.
In reality, the disk is not infinite and illuminating sources are also located at high Galactic latitudes.
However, if we adopt such realistic Milky Way model, the $g$-value should be larger than the present values to explain the derived latitude dependence at $1.25$ and $2.2\,\rm{\mu m}$ (Figure 1).
This means that the derived asymmetry factor assuming the simplest scattered light model (Jura 1979) is probably close to its lower limit.
As discussed above, the latitude dependence of additional UV-excited dust emission may be required to reduce the derived large $g$-value.

\section{SUMMARY}

Reanalyzing the DIRBE data in the near-IR bands, we first find that the intensity ratios of DGL to $100\,\rm{\mu m}$ emission increase toward low Galactic latitudes at $1.25$ and $2.2\,\rm{\mu m}$.
This trend is naturally expected from the simple scattered light model taking into account the forward scattering property of the grains, characterized by the scattering asymmetry factor $g$.
By the fitting to the derived Galactic latitude dependence, we derive the large asymmetry factor of $0.8^{+0.2}_{-0.3}$ at $1.25$ and $2.2\,\rm{\mu m}$.
The derived asymmetry factor is comparable to those obtained in the optical, but several times larger than that expected from a recent dust model.
Because the possible Galactic latitude dependence of near-IR emission from UV-excited dust would reduce the derived large asymmetry factor, the latitude dependence obtained in the present study may indicate the first detection of the additional emission component in the diffuse interstellar medium.

%% The displaymath environment will produce the same sort of equation as
%% the equation environment, except that the equation will not be numbered
%% by LaTeX.


%% Included in this acknowledgments section are examples of the
%% AASTeX hypertext markup commands. Use \url without the optional [HREF]
%% argument when you want to print the url directly in the text. Otherwise,
%% use either \url or \anchor, with the HREF as the first argument and the
%% text to be printed in the second.

\acknowledgments

We acknowledge useful discussions with K. Kawara, H. Kataza, and Y. Matsuoka.
We are grateful to the anonymous referee for a number of useful comments that greatly improved the paper.
K.S. is supported by Grant-in-Aid for Japan Society for the Promotion of Science (JSPS) Fellows. 

This publication uses the {\it COBE} datasets developed by the National Aeronautics and Space Administration (NASA) Goddard Space Flight Center under the guidance of the {\it COBE} Science Working Group.
This publication also makes use of data products from the Two Micron All Sky Survey, which is a joint project of the University of Massachusetts and the Infrared Processing and Analysis Center/California Institute of Technology, funded by NASA and the National Science Foundation.
This publication also makes use of data products from WISE, which is a joint project between the University of California, Los Angeles, and the Jet Propulsion Laboratory/California Institute of Technology.
WISE is funded by NASA.


%% in parentheses, after the keyword, but they will not be verified.

%{\it Facilities:} \facility{Nickel}, \facility{HST (STIS)}, \facility{CXO (ASIS)}.



\appendix




\clearpage

%% the file, uncomment the \plotone, \plottwo, and
%% If you need a layout that cannot be achieved with \plotone or
%% \plottwo, you can invoke the graphicx package directly with the
%% \includegraphics command or use \plotfiddle. For more information,
%% please see the tutorial on "Using Electronic Art with AASTeX" in the
%% documentation section at the AASTeX Web site, http://aastex.aas.org/
%%
%% The examples below also include sample markup for submission of
%% supplemental electronic materials. As always, be sure to check
%% the instructions to authors for the journal you are submitting to
%% for specific submissions guidelines as they vary from

%% This example uses \plotone to include an EPS file scaled to
%% 80% of its natural size with \epsscale. Its caption
%% has been written to indicate that additional figure parts will be
%% available in the electronic journal.

\clearpage

%% Here we use \plottwo to present two versions of the same figure,
%% one in black and white for print the other in RGB color
%% for online presentation. Note that the caption indicates
%% that a color version of the figure will be available online.
%%

%% This figure uses \includegraphics to scale and rotate the still frame

%% If you are not including electonic art with your submission, you may
%% mark up your captions using the \figcaption command. See the
%% No more than seven \figcaption commands are allowed per page,
%% so if you have more than seven captions, insert a \clearpage

%% Tables should be submitted one per page, so put a \clearpage before

%% Two options are available to the author for producing tables:  the
%% deluxetable environment provided by the AASTeX package or the LaTeX
%% table environment.  Use of deluxetable is preferred.
%%

%% Three table samples follow, two marked up in the deluxetable environment,
%% one marked up as a LaTeX table.

%% In this first example, note that the \tabletypesize{}
%% command has been used to reduce the font size of the table.
%% We also use the \rotate command to rotate the table to
%% landscape orientation since it is very wide even at the
%% Note also that the \label command needs to be placed

%% This table also includes a table comment indicating that the full
%% version will be available in machine-readable format in the electronic

\clearpage


\begin{thebibliography}{}
\bibitem[Arai(2015)]{arai2015} Arai, T., Matsuura, S., Bock, J., et al.\ 2015, \apj, 806, 69
\bibitem[Arendt(1998)]{arendt1998} Arendt, R.~G., Odegard, N., Weiland, J.~L., et al.\ 1998, \apj, 508, 74
\bibitem[Bernstein(2002)]{bernstein2002} Bernstein, R.~A., Freedman, W.~L., \& Madore, B.~F.\ 2002, \apj, 571, 56
\bibitem[Brandt \& Draine(2012)]{brandt2012} Brandt, T.~D., \& Draine, B.~T.\ 2012, \apj, 744, 129
\bibitem[COBEDIRBE]{cobedirbe} COBE Diffuse Infrared Background Experiment (DIRBE) Explanatory Supplement, Version 2.3. 1998, ed. M.~G. Hauser et al.
\bibitem[Compiegne(2011)]{compiegne2011} Compi\`egne, M., Verstraete, L., Jones, A., et al. \ 2011, \aap, 525, A103 
\bibitem[Cutri(2003)]{cutri2003} Cutri, R.~M., Skrutskie, M.~F., van Dyk, S., et al. 2003, The IRSA 2MASS All Sky Point Source Catalog, NASA/IPAC Infrared Science Archive
\bibitem[Draine(2011)]{draine20011} Draine, B.~T.\ 2011, Physics of the Interstellar and Intergalactic Medium (Princeton, NJ: Princeton Univ. Press)
\bibitem[Draine(2003)]{draine2003} Draine, B.~T.\ 2003, \apj, 598, 1017
\bibitem[Draine(2007)]{draine2007} Draine, B.~T., \& Li, A.\ 2007, \apj, 657, 810
\bibitem[Fitzpatrick(1999)]{fitzpatrick1999} Fitzpatrick, E.~L. \ 1999, \pasp, 111, 63
\bibitem[Henyey(1941)]{henyey1941} Henyey, L.~G., \& Greenstein, J.~L.\ 1941, \apj, 93, 70
\bibitem[Ienaka(2013)]{ienaka2013} Ienaka, N., Kawara, K., Matsuoka, Y., et al.\ 2013, \apj, 767, 80
\bibitem[Jura(1979)]{jura1979} Jura, M. \ 1979, \apj, 227, 798
\bibitem[Kelsall(1998)]{kelsall1998} Kelsall, T., Weiland, J.~L., Franz, B.~A., et al.\ 1998, \apj, 508, 44
\bibitem[Lagache(2000)]{lagache2000} Lagache, G., Haffner, L.~M., Reynolds, R.~J., \& Tufte, S.~L.\ 2000, \aap, 354, 247
\bibitem[Lehtinen(1996)]{lehtinen1996} Lehtinen, K., \& Mattila, K.\ 1996, \aap, 309, 570
\bibitem[Lehtinen(2007)]{lehtinen2007} Lehtinen, K., Juvera, M., Mattila, K., et al. \ 2007, \aap, 466, 969
\bibitem[Lillie(1976)]{lillie1976} Lillie, C.~F., \& Witt, A.~N.\ 1976, \apj, 208, 64
\bibitem[Mathis(1983)]{mathis1983} Mathis, J.~S., Mezger, P.~G., \& Panagia, N.\ 1983, \aap, 128, 212 
\bibitem[Matsumoto(2015)]{matsumoto2015} Matsumoto, T., Kim, M.~G., Pyo, J., \& Tsumura, K.\ 2015, \apj, 807, 57
\bibitem[Matsuoka(2011)]{matsuoka2011} Matsuoka, Y., Ienaka, N., Kawara, K., \& Oyabu, S.\ 2011, \apj, 736, 119 
\bibitem[Mattila(1970)]{mattila1970} Mattila, K. \ 1970, \aap, 9, 53
\bibitem[Murthy(2014)]{murthy2014} Murthy, J. \ 2014, \apjs, 213, 32
\bibitem[Sano(2015)]{sano2015} Sano, K., Kawara, K., Matsuura, S., et al.\ 2015, \apj, 811, 77 (Paper I) 
\bibitem[Sano(2016)]{sano2016} Sano, K., Kawara, K., Matsuura, S., et al.\ 2016, \apj, 818, 72 (Paper II) 
\bibitem[Schiminovich(2001)]{schiminovich2001} Schiminovich, D., Friedman, P.~G., Martin, C., \& Morrissey, P.~F.\ 2001, \apjl, 563, L161
\bibitem[Schlegel(1998)]{schlegel1998} Schlegel, D.~J., Finkbeiner, D.~P., \& Davis, M.\ 1998, \apj, 500, 525 
\bibitem[Sellgren(1992)]{sellgren1992} Sellgren, K., Werner, M.~W., \& Dinerstein, H.~L. \ 1992, \apj, 400, 238
\bibitem[Sellgren(1996)]{sellgren1996} Sellgren, K., Werner, M.~W., \& Allamandola, L.~J. \ 1996, \apjs, 102, 369 
\bibitem[Skrutskie(2006)]{skrutskie2006} Skrutskie, M.~F., Cutri, R.~M., Stiening, R.,  et al.\ 2006, \apj, 131, 1163
\bibitem[Tsumura(2013b)]{tsumura2013b} Tsumura, K., Matsumoto, T., Matsuura, S., et al.\ 2013, \pasj, 65, 120
\bibitem[Weingartner \& Draine(2001)]{weingartner2001} Weingartner, J.~C., \& Draine, B.~T.\ 2001, \apj, 548, 296 
\bibitem[Witt(1997)]{witt1997} Witt, A.~N., Friedmann, B.~C., \& Sasseen, T.~P. \ 1997, \apj, 481, 809
\bibitem[Witt(1994)]{witt1994} Witt, A.~N., Lindell, R.~S., Block, D.~L., \& Evans, R. \ 1994, \apj, 427, 227
\bibitem[Witt(2008)]{witt2008} Witt, A.~N., Mandel, S., Sell, P.~H., Dixon, T., \& Vijh, U.~P. \ 2008, \apj, 679, 497
\bibitem[Witt(1990)]{witt1990} Witt, A.~N., Oliveri, M.~V., \& Schild, R.~E.\ 1990, \aj, 99, 888
\bibitem[Wright(2010)]{wright2010} Wright, E.~L., Eisenhardt, P.~R.~M., Mainzer, A.~K., et al. \ 2010, \aj, 140, 1868
\bibitem[Zubko(2004)]{zubko2004} Zubko, V., Dwek, E., \& Arendt, R.~G.\ 2004, \apjs, 152, 211 

\end{thebibliography}
\end{document}